\documentclass{aa}  

\usepackage{graphicx}
\usepackage{txfonts}
\usepackage{url}
\usepackage{placeins}
\usepackage{float}
\begin{document}

   \title{Milky Way metallicity gradient\\ from Gaia DR2 F/1O double-mode Cepheids}
	\titlerunning{MW metallicity gradient from Gaia DR2 F/1O Cepheids}
    \authorrunning{B. Lemasle, G. Hajdu, et al.}
   \subtitle{}

   \author{B. Lemasle,   
          \inst{1}
          G. Hajdu,
          \inst{2,1,3}
          V. Kovtyukh,
          \inst{4,5}
          L. Inno,
          \inst{6,8}
          E. K. Grebel,
          \inst{1}
          M. Catelan,
          \inst{2,3}\fnmsep\thanks{On sabbatical leave at European Southern Observatory, Av. Alonso de C\'ordova 3107, 7630355 Vitacura, Santiago, Chile}\\
          G. Bono,
          \inst{7,8}
          P. Fran\c cois
          \inst{9,10}
          A. Kniazev,
          \inst{11,12.13,14}
          R. da Silva,
          \inst{8,15}
          \and
          J. Storm
          \inst{16}
          }
 
   \institute{Astronomisches Rechen-Institut, Zentrum f\"ur Astronomie der Universit\"at Heidelberg, M\"onchofstrasse 12-14, D-69120 Heidelberg, Germany
              \email{lemasle@uni-heidelberg.de}
        \and
             Instituto de Astrof\'isica, Pontificia Universidad Cat\'olica de Chile, Av. Vicu\~na Mackenna 4860, 782-0436 Macul, Santiago, Chile
		\and Instituto Milenio de Astrof\'isica, Santiago, Chile
		\and Astronomical Observatory, Odessa National University, Shevchenko Park, UA-65014 Odessa, Ukraine
        \and Isaac Newton Institute of Chile, Odessa branch, Shevchenko Park, UA-65014 Odessa, Ukraine
        \and Max-Planck-Institut f\"ur Astronomie, 69117, Heidelberg, Germany             
        \and Dipartimento di Fisica, Universit\`a di Roma Tor Vergata, via della Ricerca Scientifica 1, 00133, Rome, Italy
        \and INAF-Osservatorio Astronomico di Roma, via Frascati 33, 00078 Monte Porzio Catone, Rome, Italy
        \and GEPI, Observatoire de Paris, CNRS, Universit\'e Paris Diderot, place Jules Janssen, 92190, Meudon, France
        \and UPJV, Universit\'e de Picardie Jules Verne, 33 rue St. Leu, 80080, Amiens, France
        \and South African Astronomical Observatory, PO Box 9, 7935 Observatory, Cape Town, South Africa
        \and Southern African Large Telescope Foundation, PO Box 9, 7935 Observatory, Cape Town, South Africa        
        \and Sternberg Astronomical Institute, Lomonosov Moscow State University, Universitetskij Pr. 13, Moscow 119992, Russia        
		\and Special Astrophysical Observatory of RAS, Nizhnij Arkhyz, Karachai-Circassia 369167, Russia
        \and ASI Science Data Center, via del Politecnico snc, 00133, Rome, Italy                
        \and Leibniz-Institut f\"ur Astrophysik Potsdam, An der Sternwarte 16, D-14482 Potsdam, Germany        
             }

   \date{Received September 15, 1996; accepted March 16, 1997}

 
  \abstract
   {The ratio of the first overtone (1O) / fundamental (F) periods of mixed-mode Cepheids that pulsate simultaneously in these two modes (F/1O) is metallicity-dependent. It can therefore be used to characterize the systems that host such variable stars-
   }
   {We want to take advantage of the F/1O double-mode Cepheids listed in the Gaia DR2 catalogue to derive the metallicity gradient in the Milky Way disk.
   }
   {The metallicity is derived from the ratio of the first overtone and fundamental periods provided by Gaia DR2 while the Gaia DR2 parallaxes are used to determine the Galactocentric distances of the stars.}
   {From a visual inspection of the light curves, it turns out that a large fraction (77\%) of the Galactic F/1O double-mode Cepheids in Gaia DR2 are spurious detections. Gaia DR2 provides 3 new bona fide F/1O Cepheids. Combining them with the currently known F/1O Cepheids and using the Gaia DR2 parallaxes for the entire sample, we can derive the metallicity gradient in the Milky Way disk. We find a slope of--0.045$\pm$0.007 dex/kpc using a bootstrap method, and of --0.040$\pm$0.002 dex/kpc using a total least squares method. These results are in good agreement with previous determinations of the [Fe/H] gradient in the disk based on canonical Cepheids.
   }
   {The period ratio of F/1O Cepheids allows for a reliable determination of the metallicity gradient in the Milky Way, and in turn, in other systems that would be difficult to reach via classical spectroscopic methods.}


   \maketitle
%

\section{Introduction}

\par Cepheids are pulsating variable stars and the vast majority of them pulsates in a single mode, in general either the fundamental (F), or the first-overtone (1O), or the second-overtone (2O) mode. A small fraction of the Cepheids pulsates in two modes simultaneously (in general F and 1O or 1O and 2O). A few rare objects even pulsate in three modes simultaneously \citep[e.g.,][]{Moskalik2014,Poretti2014,Soszynski2015b}.

\par Double-mode Cepheids have been used to test the stellar evolutionary and pulsation models \citep[e.g.,][]{Buchler2007,Buchler2008,Smolec2010}. They have also been used to study the stellar populations of the few galaxies in which they have been discovered, namely the Milky Way \citep{Oosterhoff1957a,Oosterhoff1957b}, M31 \citep{Poleski2013,Lee2013}, M33 \citep{Beaulieu2006}, the Large Magellanic Cloud (LMC) \citep[e.g.,][]{Alcock1999,Soszynski2008a,Marquette2009} and the Small Magellanic Cloud (SMC) \citep[e.g.,][]{Marquette2009,Soszynski2010}.

\par Petersen diagrams \citep{Petersen1973} where the period ratios of mixed-mode Cepheids are plotted versus the longer period are very useful tools to study the properties of these stars. 
Combined with the huge amount of data provided by the OGLE microlensing survey \citep{Udalski2015}, they led to numerous discoveries in recent years for both Cepheids and RR Lyrae stars \citep[e.g.,][]{Coppola2015,Smolec2016,Prudil2017}.

\par In Petersen diagrams, it has been known for a long time that the period ratios fall around $P_{21}= P_{2}/P_{1}=0.80$ for the Cepheids pulsating simultaneously in the first and second overtone modes (1O/2O), and around $P_{10}= P_{1}/P_{0}=0.72$ for the Cepheids pulsating simultaneously in the fundamental and first overtone modes (F/1O). Moreover, the period-ratios are different for the F/1O double-mode Cepheids located in the Milky Way, the LMC and the SMC, and these stars occupy different regions of the Petersen diagram, which is a consequence of the metallicity difference between these galaxies \citep[e.g.,][]{Buchler2007}. In contrast, $P_{21}$ is metallicity-independent. Several authors \citep{Sziladi2007,Kovtyukh2016} have calibrated the relation between $P_{1}/P_{0}$ and [Fe/H] using high-resolution spectroscopy of Galactic Cepheids and used it to study for instance the metallicity distribution of the young population in the Magellanic Clouds.

\par In this paper, we want to apply this same calibration relation to the F/1O double-mode Cepheids newly discovered by Gaia \citep{GaiaCollaboration2016,GaiaCollaboration2018} in order to derive their metallicities. Combining it with accurate distances determined directly from the Gaia DR2 parallaxes or using period-luminosity relations, we can derive the metallicity gradient in the Galactic disk. Such gradients (and their temporal evolution) provide strong constraints on the mechanisms driving the chemo-dynamical evolution of the Milky Way \citep[e.g.,][]{Minchev2018,Navarro2018,Prantzos2018,Grisoni2018} and on the relative importance of e.g., stellar radial migration or the possible variation of the star formation efficiency.

\par The paper is organized as follows: in Sect.~\ref{current} we briefly describe the sample of currently known F/1O double-mode Cepheids in the Milky Way while in Sect.~\ref{new} we analyze the new F/1O candidates in Gaia DR2. In Sect.~\ref{Others} we comment on individual variable stars. Sect.~\ref{grad} is dedicated to the determination of the Milky Way metallicity gradient. Results are summarized in Sect.~\ref{Conc}.

\section{F/1O double-mode Cepheids currently known in the Milky Way}
\label{current}

\par Only 27 F/1O Cepheids are known in the Milky Way disk\footnote{A similar number of 1O/2O double-mode Cepheids (that are not relevant for this paper) have also been discovered: CO Aur and V1048 Cen have been known for a long time; V2157 Sgr, V767 Sgr, V363 Cas have been reported by \cite{Hajdu2009a}, and several more stars have been identified by e.g., \cite{Soszynski2011,Pietrukowicz2013,Khruslov2013,Khruslov2016} in large scale photometric surveys.}. The coordinates and properties of those known for a long time are listed in the McMaster Cepheid database\footnote{\url{https://www.physics.mcmaster.ca/Cepheid/BeatCepheid.html}}, to which we added V901 Mon \citep{Antipin2006}. A few more have been recently reported in the disk, the bulge or the far side of the disk by the OGLE survey \citep{Soszynski2011,Pietrukowicz2013}. \cite{Kovtyukh2016} also reported the chemical composition for 18 of them. We gathered all the relevant information for the sample used in this paper in Table~\ref{knownMWF/1O}.

\begin{table*}[!htp]
\caption{F/1O double-mode Cepheids currently known in the Milky Way. $P_{1}/P_{0}$ are from the McMaster Cepheid database, except for V371 Per, from \cite{Wils2010}, V901 Mon from \citet{Antipin2006}, and for the OGLE Cepheids, from \cite{Soszynski2011}, and \cite{Pietrukowicz2013}. When available, metallicities are from \cite{Kovtyukh2016}.}             
\label{knownMWF/1O}
\centering
\begin{tabular}{l r@{\hskip6pt}l l r r c r r} 
\hline\hline
\multicolumn3c{Name} & \multicolumn1c{RA (J2000)} & \multicolumn1c{Dec (J2000)} & \multicolumn1c{$P_{0}$} & $P_{1}/P_{0}$ & [Fe/H] & \multicolumn1c{Gaia DR2 ID} \\
\multicolumn3c{ } & \multicolumn1c{hms} & \multicolumn1c{dms} & \multicolumn1c{d} & \multicolumn1c{} & \multicolumn1c{dex} & \multicolumn1c{} \\
\hline  
& V825 & Cas & 00 25 18.17 & +60 45 53.3  & 3.7342 & 0.7103 &       &  428835686898545920 \\
&   AS & Cas & 00 25 37.72 & +64 13 47.6  & 3.0247 & 0.7127 & -0.19 &  431184518613946112 \\  
&   TU & Cas & 00 26 19.45 & +51 16 49.3  & 2.1393 & 0.7097 &  0.04 &  394818721274314112 \\  
& V371 & Per & 02 55 31.19 & +42 35 19.8  & 1.7371 & 0.7312 & -0.40 &  336558933011470720 \\
& V901 & Mon & 06 27 25.25 & +01 11 32.4  & 2.26571 & 0.7124 &       &  3123433228497249280 \\  
&   DZ & CMa & 07 16 59.33 & -15 18 25.4  & 2.3629 & 0.7195 &       & 3031419601502352768 \\ 
&   VX & Pup & 07 32 36.65 & -21 55 49.5  & 3.0109 & 0.7104 & -0.08 & 5619192029327633664 \\  
&   BE & Pup & 07 33 35.50 & -25 50 37.2  & 2.87 & 0.7136 &       & 5613375028701902976 \\  
&   AX & Vel & 08 10 49.32 & -47 41 54.8  & 3.6732 & 0.7059 & -0.05 & 5519196703818908800 \\ 
&   AP & Vel & 08 39 45.76 & -43 51 39.2  & 3.1278 & 0.7033 &  0.07 & 5523256203825403392 \\   
& V701 & Car & 10 09 13.62 & -57 14 33.4  & 4.089 & 0.7017 &  0.06 & 5258904608894451072 \\  
&   GZ & Car & 10 20 20.37 & -59 22 35.8  & 4.1589 & 0.7054 &  0.02 & 5255066591776722432 \\  
&    Y & Car & 10 33 10.85 & -58 29 55.1  & 3.6398 & 0.7032 &  0.03 & 5351428787262634624 \\  
&   EY & Car & 10 42 23.03 & -61 09 57.3  & 2.876 & 0.7079 &  0.04 & 5254070090673438592 \\  
&   UZ & Cen & 11 40 58.54 & -62 41 32.9  & 3.3344 & 0.7064 & -0.03 & 5333340824575196800 \\  
&   BK & Cen & 11 49 16.02 & -63 04 42.9  & 3.1739 & 0.7004 &  0.13 & 5333259323269569792 \\  
&V1210 & Cen & 14 36 55.56 & -58 15 41.2  & 4.317 & 0.7035 &  0.03 & 5891313563729348480 \\ 
&    U & TrA & 16 07 19.00 & -62 54 38.0  & 2.5684 & 0.7105 & -0.09 & 5829232354047904256 \\   
& V458 & Sct & 18 22 27.07 & -10 07 29.2  & 4.84125 & 0.6993 &  0.11 & 4154536747505387648 \\  
& V367 & Sct & 18 33 35.24 & -10 25 38.0  & 6.2931 & 0.6967 &  0.07 & 4155020566971108224 \\  
&   BQ & Ser & 18 36 15.94 & +04 23 53.7  & 4.2707 & 0.7053 & -0.05 & 4283775646336574336 \\   
&   EW & Sct & 18 37 51.11 & -06 47 48.5  & 5.8232 & 0.6985 &  0.04 & 4253017873709602304 \\
\multicolumn3l{OGLE-BLG-CEP-03\footnote{This Cepheid is located on the far side of the Galactic disk, in a flared outer disk, according to \cite{Feast2014}.}} & 17 44 43.79 & -23 43 25.1 & 1.2356978 & 0.7327 &    & 4068367505901854080 \\
\multicolumn3l{OGLE-BLG-CEP-21}  & 17 57 50.37 & -28 04 43.3 & 0.7785577 & 0.7334 &    & 4062757346569515520 \\
\multicolumn3l{OGLE-GD-CEP-0009} & 10 58 58.43 & -61 52 18.3 & 1.676337 & 0.7246 &    & 5241828295658352000 \\
\multicolumn3l{OGLE-GD-CEP-0012} & 11 04 58.71 & -62 01 52.4 & 0.6557404 & 0.7689 &    & 5337135960744652800 \\
\multicolumn3l{OGLE-GD-CEP-0016} & 13 22 55.06 & -65 00 03.6 & 2.649648 & 0.7408 &    & 5858806880457739776 \\
\hline
\end{tabular}
\end{table*}


\section{New Galactic F/1O double-mode Cepheids in Gaia DR2}
\label{new}

\subsection{F/1O double-mode Cepheids in Gaia DR2}

\par We searched for the F/1O double mode Cepheids in Gaia DR2, in the dedicated catalogue for Cepheids (\texttt{gaiadr2.vari\_cepheid}) that contains 9572 Cepheid candidates \citep{Clementini2018}. We found 162 stars for which the keyword \texttt{mode\_best\_classification} indicates "\texttt{MULTI}" and the \texttt{multi\_mode\_best\_classification} indicates "\texttt{F/1O}".

\par However, a large fraction ($>$80\%) of them are located in the Magellanic Clouds. To date, many more double-mode Cepheids have been discovered (mostly via microlensing surveys) in the LMC/SMC than in the Milky Way. For instance, OGLE \citep{Udalski2015} reports respectively 95/68 F/1O and 322/239 1O/2O double-mode Cepheids in the LMC/SMC. The difference between the Milky Way and the Magellanic Clouds concerning double-mode Cepheids is caused by the mean metal content affecting the topology of the instability strip \citep{Bono2002d}.

\par Since we are not interested in the LMC/SMC Cepheids in our study of the Galactic metallicity gradient, it is an easy task to select them out (based on their position on the sky) in order to restrict our sample to the candidate double-mode Cepheids located in the Milky Way. We are then left with 30 stars. Their properties are listed in Table~\ref{comp}.

\par 
\begin{table*}[!ht]
\scriptsize
\caption{Properties of candidate Galactic F/1O double-mode Cepheids in Gaia DR2. Genuine F/1O Cepheids (See Sect.~\ref{butterflies}) are marked in bold. Note that the secondary period of Gaia DR2 5845572265108049408 has been revised.}        
\label{comp}
\centering
\begin{tabular}{r r r r c c c r c} 
\hline\hline
\multicolumn1c{ Gaia DR2 source ID}   &  \multicolumn1c{RA (J2015.5)}  &  \multicolumn1c{Dec (J2015.5)}  & \multicolumn1c{$G$}    & $P_{0}$        & $P_{1}$        & $P_{1}/P_{0}$  &\multicolumn1c{parallax}           & parallax error\\ 
\multicolumn1c{}   &  \multicolumn1c{deg}  &  \multicolumn1c{deg}  & \multicolumn1c{mag}    & d        & d &  & \multicolumn1c{mas}           & mas\\ 
\hline  
514736269771300224  &  33.8810955 &  63.5177809  & 10.318 &  4.395504 & 3.298857 & 0.7505 &  0.16559 & 0.02877 \\ 
466906311366699520  &  48.4458726 &  63.3494795  & 13.184 &  3.032677 & 2.182606 & 0.7197 &  0.11580 & 0.02284 \\  
462252662762965120  &  50.9495124 &  59.3556690  & 11.843 &  4.156364 & 2.949602 & 0.7097 &  0.22239 & 0.03658 \\
3103637208835609728 &  99.2347677 &  -5.3509912  & 11.388 &  5.028469 & 3.647813 & 0.7254 &  2.39838 & 0.03595 \\
5593427031607304704 & 112.1240227 & -30.6555461  & 12.052 &  3.305151 & 2.370890 & 0.7173 &  0.02406 & 0.02410 \\
5613375028701902976 & 113.3979224 & -25.8436670  & 13.271 &  2.872002 & 2.049570 & 0.7136 &  0.15595 & 0.02354 \\  
5599566983722741248 & 114.4770586 & -29.4380410  & 14.182 &  3.583169 & 2.620036 & 0.7312 &  0.08720 & 0.02232 \\
3036328405518444800 & 118.9644056 & -12.0286246  & 16.475 &  5.174338 & 3.917119 & 0.7570 &  0.07228 & 0.07333 \\  
3068089482512577152 & 119.7726701 &  -5.6955677  & 14.276 &  6.802241 & 5.055228 & 0.7432 &  0.81402 & 0.02881 \\ 
5594100246268225280 & 119.8429790 & -33.3571287  & 13.393 &  7.301897 & 5.281963 & 0.7234 &  0.03920 & 0.01976 \\ 
5519196703818908800 & 122.7054723 & -47.6985287  &  7.990 &  3.672245 & 2.592817 & 0.7061 &  0.61614 & 0.03087 \\
5431347477101421824 & 146.1401577 & -41.5621957  & 14.007 &  3.248157 & 2.400509 & 0.7390 &  0.92039 & 0.01827 \\ 
5254070090673438592 & 160.5959094 & -61.1659013  & 10.022 &  2.876475 & 2.034831 & 0.7074 &  0.33047 & 0.02535 \\
{\bf 5369956245371775104} & 174.0420646 & -50.1058022  & 14.621 &  3.172059 & 2.256076 & 0.7112 &  0.73799 & 0.02800 \\
5333340824575196800 & 175.2438449 & -62.6924746  &  8.535 &  3.334994 & 2.355877 & 0.7064 &  0.62136 & 0.03000 \\  
{\bf 5845572265108049408} & 201.9938677 & -67.4170982  & 14.621 &  1.086558 & 0.796466 & 0.7330 &  0.13955 & 0.02179 \\
6117651360865355136 & 213.4847207 & -38.0963457  &  8.394 &  5.029705 & 3.745449 & 0.7447 &  0.44801 & 0.06237 \\  
5895841249526120064 & 215.8222650 & -54.9598377  & 16.080 &  3.158062 & 2.312072 & 0.7321 &  0.77369 & 0.08533 \\
1693501722163309312 & 226.5083236 &  67.4746206  & 18.868 &  1.035847 & 0.763668 & 0.7372 &  0.06442 & 0.17579 \\
4346080262981428224 & 239.0989956 & -11.3102248  & 14.721 &  3.417113 & 2.527238 & 0.7396 &  1.00726 & 0.03785 \\   
4549519051176647808 & 265.4641799 &  16.8317200  & 14.412 &  6.058693 & 4.331957 & 0.7150 &  0.15405 & 0.02313 \\
4578235236881587968 & 272.6974491 &  23.2404164  & 17.713 &  5.790263 & 4.253632 & 0.7346 &  0.22153 & 0.09585 \\ 
4038015379997952512 & 272.9976567 & -36.1112230  & 11.798 &  1.014954 & 0.761395 & 0.7502 &  0.15719 & 0.04333 \\ 
4265371574109405824 & 284.3511067 &  -0.7302387  & 10.560 &  4.182775 & 2.988056 & 0.7144 &  0.39987 & 0.04251 \\  
6710614339593008384 & 284.7512563 & -46.4247479  & 16.066 &  0.920853 & 0.683047 & 0.7418 &  0.04556 & 0.07073 \\ 
4221891970813502464 & 300.6769693 &  -3.6096227  & 14.141 &  1.210175 & 0.874649 & 0.7227 &  0.10363 & 0.03285 \\ 
1823617898156364160 & 301.5620587 &  20.7362011  & 17.084 &  1.139916 & 0.830942 & 0.7290 & -0.15773 & 0.09344 \\ 
1807821313362785664 & 302.2597270 &  15.7689687  & 17.209 &  4.091437 & 3.064627 & 0.7490 &  0.46988 & 0.09362 \\ 
{\bf 2166389269407411200} & 312.8661427 &  46.3035145  & 12.337 &  3.162273 & 2.237166 & 0.7075 &  0.12807 & 0.02847 \\ 
6394890542044709888 & 342.3867034 & -61.8731813  & 15.209 &  5.492661 & 4.054042 & 0.7381 &  0.71988 & 0.03292 \\
\hline
\end{tabular}
\end{table*}

\par By cross-matching these Cepheids with the 27 currently known Galactic Cepheids, we find that only four of them have been recovered by Gaia as F/1O Cepheids, namely UZ Cen, AX Vel, EY Car, BE Pup. In Table~ \ref{recovered}, we list the periods and period ratios provided by Gaia and by the McMaster database. The periods and the period ratios are in excellent agreement.

\begin{table*}[!ht]
\caption{Variability properties of double-mode Cepheids in common in the McMaster database (left panel) and Gaia DR2 (right panel).}        
\label{recovered}
\centering
\begin{tabular}{c c c c | c c c c c} 
\hline\hline
ID & $P_{0}$ & $P_{1}$ & $P_{1}/P_{0}$ & Gaia source ID & $P_{0}$ & $P_{1}$ & $P_{1}/P_{0}$ \\
   &    d    &    d    &               &                &              d   &    d   &               \\
\hline
UZ Cen & 3.3344 & 2.3553 & 0.7064 & 5333340824575196800 & 3.33499416 & 2.35587694 & 0.7064 \\
AX Vel & 3.6732 & 2.5929 & 0.7059 & 5519196703818908800 & 3.67224527 & 2.59281694 & 0.7061\\
EY Car & 2.876 & 2.036 & 0.7079 & 5254070090673438592 & 2.87647482 & 2.03483108 & 0.7074\\
BE Pup & 2.870 & 2.048 & 0.7136 & 5613375028701902976 & 2.87200198 & 2.04957028 & 0.7136\\
\hline
\end{tabular}
\end{table*}
	 	
\par In addition to the 4 stars quoted above, 7 known double-mode Cepheids are identified in the \texttt{gaiadr2.vari\_cepheid} catalogue, but as single-mode pulsators, either in the fundamental (U Tra, BK Cen, GZ Car, V458 Sct, TU Cas) or in the first-overtone mode (Y Car, V825 Cas).

\par In order to clarify the status of the 16 remaining stars, we first checked the main Gaia catalogue, which indicates that three of the known double-mode Cepheids are not recognized as variable stars, namely V367 Sct, V371 Per, and V901 Mon. Then we also investigated the \texttt{gaiadr2.vari\_classifier\_result} catalogue \citep{Holl2018}. With the exception of the three stars reported as non-variable, all the double-mode Cepheids that have long been known can be found in this catalogue, 15 as classical Cepheids and 4 as type II Cepheids (AS Cas, DZ CMa, V825 Cas and BE Pup). It is not surprising that the \texttt{gaiadr2.vari\_cepheid} catalogue contains only a fraction of the Cepheids identified by the variability classifier, and in particular that the stars for which fewer epochs were observed are missing. We note that V825 Cas, classified as a type II Cepheid (but with a \texttt{best\_class\_score} close to 0) by the \texttt{multi-stage random forest semi-supervised classifier}, is tabulated as a first-overtone classical Cepheid in the \texttt{gaiadr2.vari\_cepheid} catalogue. Also BE Pup, identified as a type II Cepheid (\texttt{best\_class\_score}$\approx$ 0.34) by the same classifier, is properly listed as a double-mode Cepheid in the \texttt{gaiadr2.vari\_cepheid} catalogue. The five F/1O Cepheids recently discovered by OGLE are not present in the \texttt{gaiadr2.vari\_cepheid} or in the \texttt{gaiadr2.vari\_classifier\_result} catalogue.

\subsection{Visual inspection of F/1O Cepheid candidates} 

\par Given the discrepancies we decided to check the 26 (30-4) new F/1O Cepheids listed in Gaia DR2. We searched the Simbad and AAVSO VSX databases \citep{Watson2006} for information on the F/1O Cepheid candidates. In a few cases, stars were already classified as non-Cepheid variables in different works. In clear-cut cases, we have accepted the classification provided by the authors of these works, as indicated by the comments we give upon individual variables in Sect~\ref{Others}.

\par A number of stars have been previously classified as single-mode Cepheids, while the rest were previously unknown variables. We have visually inspected the Gaia DR2 $G$-band light curves of each of these variables. For some, a simple inspection of the light curve folded with the main periodicity was enough to establish the single-mode nature of these stars. For these, the limited amount of data points, together with the light curve gaps and the automated classification procedures must have led to the wrongful claim of a secondary periodicity.

\par
\begin{figure*}[h]
\centering
\includegraphics[width=17cm]{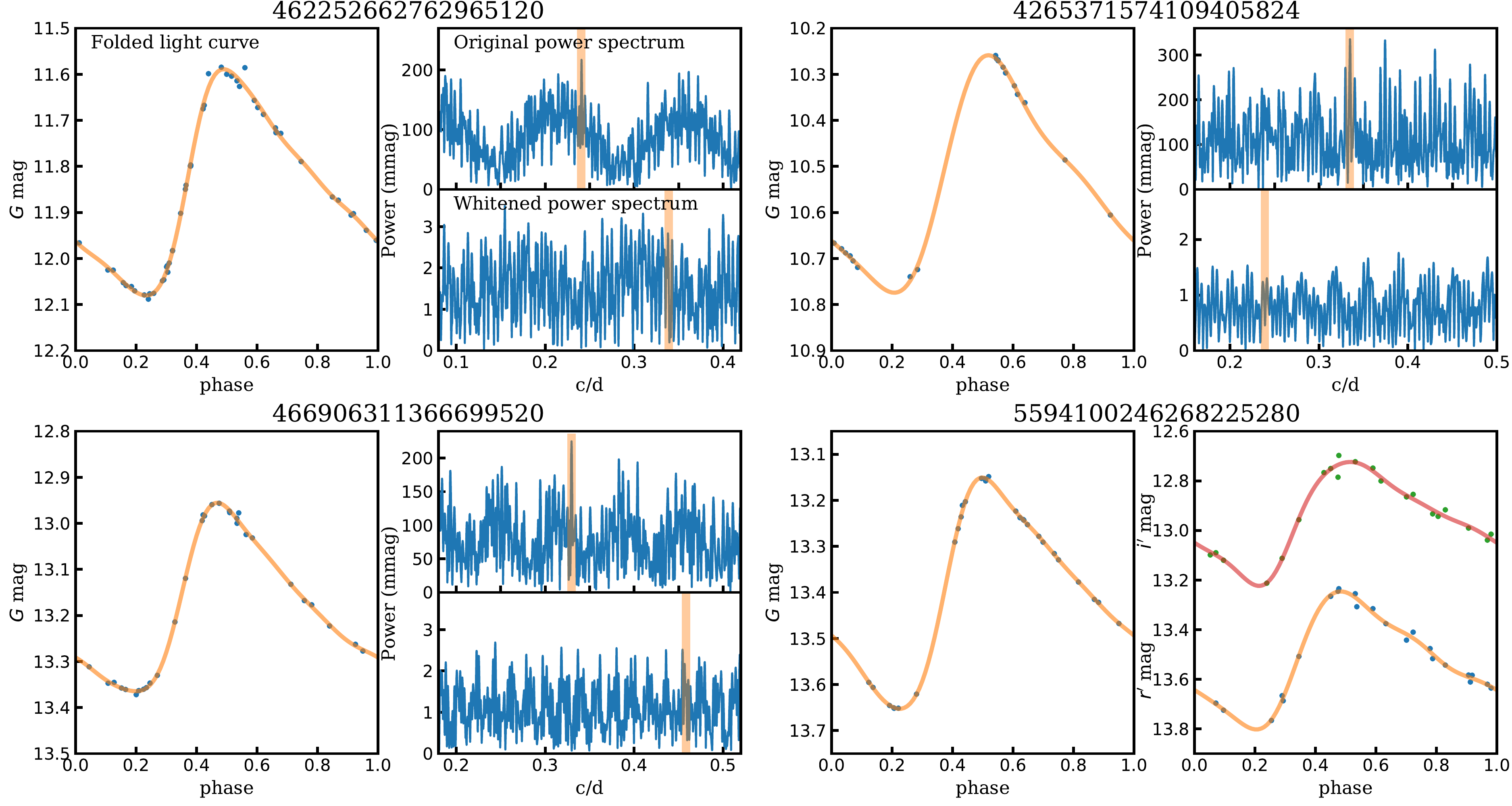}
\caption{Example light curves and power spectra of selected variables from our analysis of the Gaia DR2 F/1O double-mode Cepheid sample. For each of the four stars, the left-hand panels show the Gaia G-band light curves folded with the periodicity found by our analysis. For three stars, the smaller panels show their original (top) and whitened (bottom) power spectra in the frequency region where the fundamental and first-overtone signals are expected. On the top panels, vertical bars are showing the identified periodicity of each star. Meanwhile, on the bottom panels, the vertical bars are showing the location of the secondary periodicity given by the Gaia DR2 variability catalog \citep[\texttt{gaiadr2.vari\_cepheid}][]{Clementini2018}. For the last star, the right-hand panel shows light curves available for the variable in the literature \citep{Hackstein2015}, which reaffirm our analysis of the Gaia light curves.}
\label{examples}
\end{figure*}

 The Gaia DR2 G-band light curves of the rest of the F/1O Cepheid candidates were inspected, using variability analysis based on the Discrete Fourier Transform \citep{Deeming1975}, as implemented in the package \textsf{Mufran} \citep{Kollath1990} and the non-linear harmonic fitting routine \textsf{lcfit}. The main periodicities were identified in the discrete Fourier spectra of the successively pre-whitened (the so-far identified frequencies removed) light curves of the stars. For a number of variables, we have inspected photometry from other sources, mostly from the ASAS-SN survey \citep{Shappee2014,Kochanek2017} to confirm the periods appearing in the DR2 data. Examples of such analyses are shown in Fig.~\ref{examples}.

\par We have found that only three of the 26 new Galactic F/1O Cepheid candidates in Gaia DR2 can be really classified as such based on previous evidence and our analysis. These variables are listed in Sect.~\ref{butterflies}. For the rest of the 23 variables, we give our reasons for discarding them as F/1O Cepheids in Sect.~\ref{Others}.

\subsection{Efficacy of F/1O Cepheid identification in Gaia DR2}

\par Our analysis allows us to revise the efficiency of the variability analysis pipeline devised for the classification of double-mode Cepheids, when applied to the Gaia DR2 data. As mentioned before, four of the 30 F/1O Cepheid candidates were already known before, and three additional stars turned out to be real F/1O Cepheids, according to our analysis. Therefore, for the specific case of F/1O Cepheids, the precision of the Gaia DR2 sample in the Galactic field is $7/30 \sim 23 \%$. Furthermore, as mentioned before, 27 F/1O Cepheids have been known in the Galactic field, and this current data release has only recovered four of them, therefore the recall is $4/27 \sim 15 \%$. 
It is expected that these numbers will improve with the number of data points in the light curves in forthcoming Gaia releases. 

\subsection{The Petersen diagram} 

\par The Petersen diagram \citep{Petersen1973} is a useful tool to exploit the information on the physical properties of pulsating variable stars that pulsate in at least two modes simultaneously. It shows the period ratio $P_{S}/P_{L}$ of a shorter period $P_{S}$ and a longer period $P_{L}$ as a function of $P_{L}$ or  $\log P_{L}$. Observed periods are usually estimated with good accuracy and the error bars in this plot are so small that they remain invisible.

\par
\begin{figure*}[!ht]
\centering
\includegraphics[width=8.5cm]{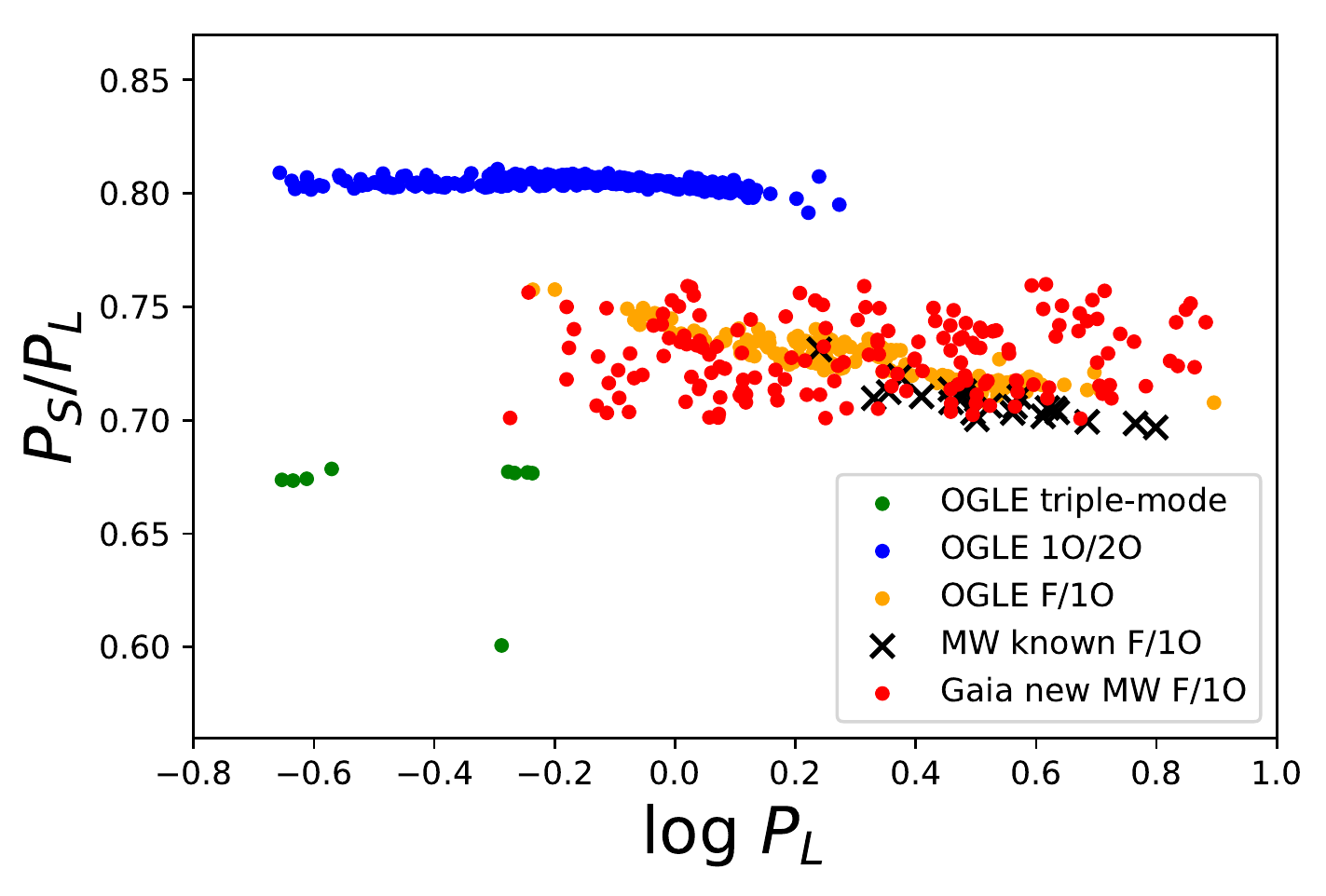}
\includegraphics[width=8.5cm]{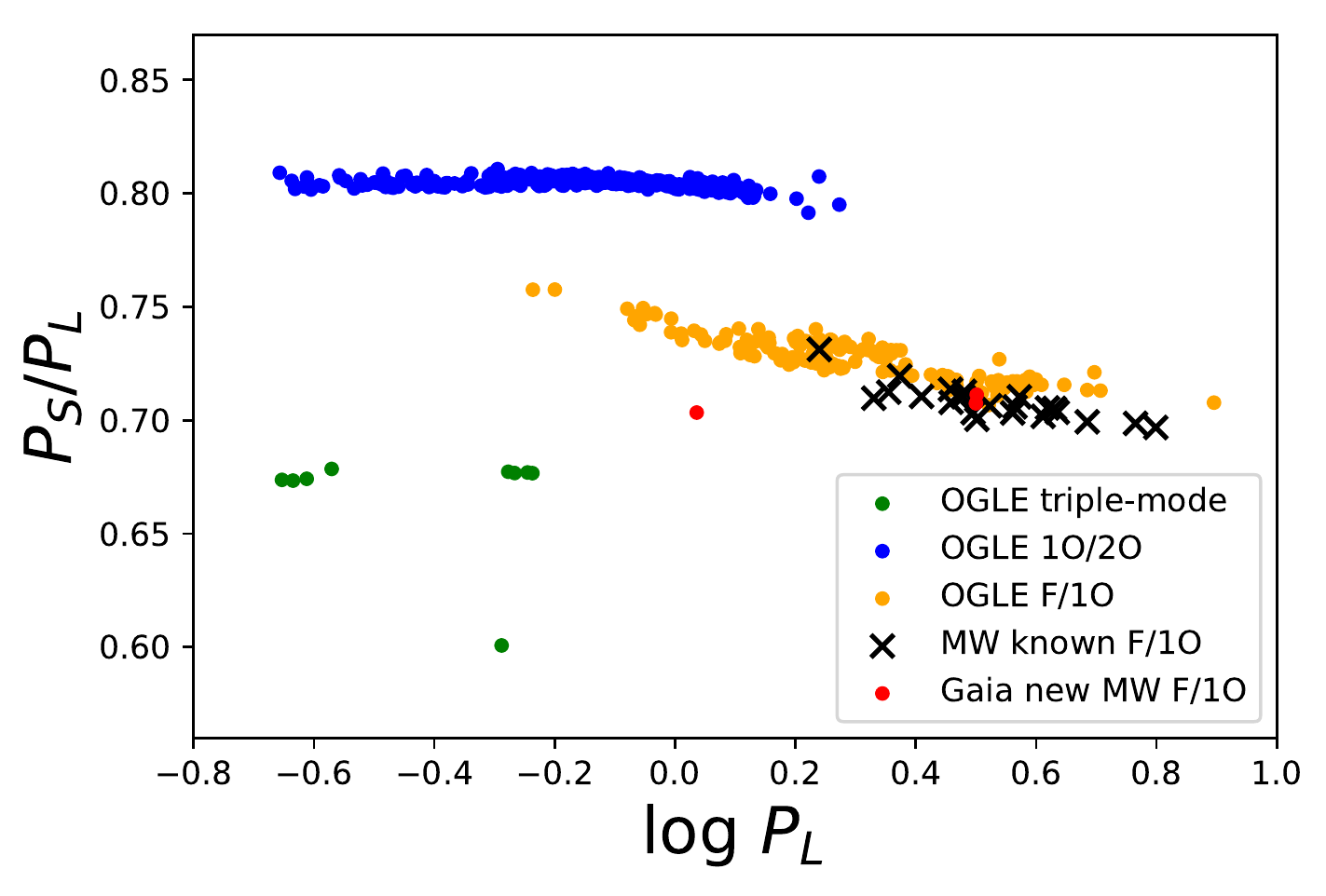}
\caption{Petersen diagrams (see text for explanations).
{\it Left:} The entire Gaia DR2 sample of F/1O Cepheids (including those located in the Magellanic Clouds, 162 stars in total) is displayed over OGLE data for different classes of Cepheids pulsating in different modes. Previously known F/1O Cepheids in the Milky Way are shown as black crosses.
{\it Right:} The bona fide Gaia DR2 sample of F/1O Cepheids (3 stars) is displayed over OGLE data for different classes of Cepheids pulsating in different modes. Previously known F/1O Cepheids in the Milky Way are shown as black crosses.}
\label{Petersen}
\end{figure*}

 It was clear from the very beginning that the Gaia DR2 sample of F/1O double-mode Cepheids (including the Magellanic ones) was highly contaminated, since the majority of the candidates reported fall outside the locus of F/1O double-mode Cepheids in the Petersen diagram mostly based on OGLE IV data. (See Fig.~\ref{Petersen}, left panel). This can arise either from a misclassification of the star or from a wrong determination of the period(s).

\par In Fig.~\ref{Petersen} (right panel), we show the Petersen diagram including only the confirmed Galactic candidates. Two out of the three new F/1O Cepheids fall well within the locus occupied by the Milky Way F/1O Cepheids.

\subsection{New Galactic F/1O Cepheids from Gaia DR2}
\label{butterflies}

\noindent{\it 2166389269407411200}: This variable is V1533~Cyg, listed by Simbad as an RR~Lyrae variable, based on the classification of \cite{Hoffman2009}. However, \cite{Wils2004} list this star as a Cepheid. Our analysis of the 
ASAS-SN data supports the latter classification as a Cepheid variable, pulsating simultaneously in the fundamental and first-overtone modes with the periods given by Gaia DR2. \newline
\noindent{\it 5369956245371775104}: Our analysis confirms that this star is a real F/1O double-mode Cepheid.\newline
\noindent{\it 5845572265108049408}: Analyses of the Gaia light curve have revealed that the Gaia DR2 period given for the first overtone, 0.79647\,days is an alias of the real secondary period, 0.76426\,days. The ASAS-SN light curve confirms our finding, resulting in a decrease of the period ratio from 0.7330 to 0.7034, as well.\newline
In the following, these stars are sometimes identified (for convenience reasons) as Gaia1, Gaia2, and Gaia3, respectively.

\section{Stars misclassified as Galactic F/1O Cepheids}
\label{Others}

\noindent{\it 462252662762965120}: This variable is AC~Cam, a known fundamental-mode classical Cepheid. Analysis of the Gaia light curve does not provide convincing evidence for a secondary pulsation mode, as illustrated by the top left panels of Fig.~\ref{examples}.\newline
\noindent{\it 466906311366699520}: This star, also known as ASASSN-V J031346.98+632052.8, has been reported to be a fundamental-mode classical Cepheid by \cite{Jayasinghe2018}.
Likewise, the Gaia light curve does not show evidence of a secondary periodicity, as shown by the bottom left panels of Fig.~\ref{examples}.\newline
\noindent{\it 514736269771300224}: We have found no evidence of additional periodicities in the Gaia light curve besides the claimed first overtone period. Inspection of the ASAS-SN light curve supports this result, and the light curve shape suggests the classification of this variable as a first-overtone Cepheid.\newline
\noindent{\it 1693501722163309312}: This variable is associated with the Ursa~Minor dwarf galaxy, with a membership probability of 0.96 \citep[][ID 296 in their Table 1]{Eskridge2001}.
Its position above the horizontal branch (HB), and relatively long pulsation period of $\sim 0.7637$\,days,
suggest that this object is a post-ZAHB star crossing the instability strip. There is no sign of additional periodicities in the light curve.\newline
\noindent{\it 1807821313362785664}: Inspection of the Gaia light curve reveals that for the main periodicity of $\sim 4.091437$\,days, the entirety of the rising branch is missing, which, coupled with the relatively high scatter due to the faintness of the star, has led to the incorrect classification of this variable as an F/1O Cepheid.\newline
\noindent{\it 1823617898156364160}: There is no sign of an additional mode in the Gaia light curve of this variable besides the fundamental mode. Its similar period and light curve shape suggests the same variability type as 4221891970813502464.\newline
\noindent{\it 3036328405518444800}: Analysis of the Gaia light curve does not provide evidence for the claimed secondary period. The light curve is missing the rising branch, when folded with the main period, which together with the small amplitude ($\sim0.3$\,mag) prevents the identification of the variable class of this star.\newline
\noindent{\it 3068089482512577152}: Neither the Gaia, nor the ASAS-SN light curves of this star show any sign of the claimed secondary period. Due to the relatively low total amplitude ($\sim 0.35$ mag in G) and the phase gap on the descending branch of the light curve, the type of this variable is uncertain.\newline
\noindent{\it 3103637208835609728}: This candidate Cepheid is the active star ASAS~J063656-0521.0, analyzed in detail by \cite{Savanov2014}. We note that active/spotted stars have been known to be masquerading as pulsating variables, when the length of photometric observations and/or the methods of analysis do not allow for a clear distinction \citep[e.g.,][]{Pietrukowicz2015a}.\newline
\noindent{\it 4038015379997952512}: This star is the peculiar variable V725~Sgr. \cite{Swope1937} reported this star to be pulsating with a period that increased from 12\,days in 1926, to 21\,days in 1935. Since then, the star has become a semi-regular variable with a period of $\sim90$\,days \citep{Percy2006}, though even more recently it has also been reported to be an irregular variable \citep{Battinelli2010}. The low number of Gaia $G$ epochs (18) led to the misclassification of this variable as a F/1O Cepheid.\newline
\noindent{\it 4221891970813502464}: This star has been classified as an AHB1 / XX~Vir variable\footnote{For the relation between AHB, XX~Vir, BL~Her and type~II Cepheids more generally, the reader is referred to Sect.~7.2 in \cite{Catelan2015}.
} \citep{Sandage2006a} by \cite{Jayasinghe2018}, based on photometry from the ASASS-N survey, under the name ASASSN-V J200242.48-033634.4. The light curve is monoperiodic without any additional mode.\newline
\noindent{\it 4265371574109405824}: This variable is the fundamental-mode classical Cepheid V493~Aql. There is no hint of a secondary periodicity in the light curve, but the inspection of the folded Gaia light curve on Fig.~\ref{examples} reveals phase gaps, which probably led to the incorrect classification of this variable.\newline
\noindent{\it 4346080262981428224}: This variable was classified as a binary under the name CSS\_J155623.7-111836 by \cite{Drake2014}. Our analysis of the Gaia light curve supports the claimed fundamental period of $\sim 3.41711$\,days, but we find no evidence of a secondary periodicity. Based on the light curve shape, this variable is most probably of the BL~Her type.
\newline
\noindent{\it 4549519051176647808}: This star is likely a binary (probably of EA (Algol) type), with twice the period given for the fundamental mode in the Gaia variability catalog, $\sim 12.117$\,days. \newline
\noindent{\it 4578235236881587968}: This star does not show a secondary period in its light curve. The relatively high scatter due to its faintness, as well as its total variability amplitude of only $\sim0.35$\,mag do not permit an unambiguous classification of variability type. \newline
\noindent{\it 5431347477101421824}: The light curve shows a major gap on the rising branch, when folded with the period of the claimed fundamental mode. Analysis of the ASAS-SN light curve does not show the claimed first overtone period, but reveals changes in the mean magnitude of the variable on a yearly time scale, indicating that this star is probably an active star like 3103637208835609728, despite being classified as a Cepheid by \cite{Jayasinghe2018}.\newline
\noindent{\it 5593427031607304704}: The analysis of the Gaia and ASAS-SN light curves did not reveal any sign of the additional periodicity. The light curve shape indicates that this variable is a first-overtone classical Cepheid.\newline
\noindent{\it 5594100246268225280}: Inspection of the Gaia light curve folded with the period given for the first overtone mode unambiguously reveals the single, fundamental mode nature of this variable. We note that the $r'$ and $i'$-band photometry published by \cite{Hackstein2015} (for source GDS\_J0759223-332125) also supports that classification of this variable, as illustrated by the bottom right-hand panels of Fig.~\ref{examples}.\newline
\noindent{\it 5599566983722741248}: This variable has been classified as a fundamental-mode classical Cepheid by \cite{Jayasinghe2018}. The Gaia light curve presents significant phase gaps when folded with the pulsation period, probably leading to the misclassification of this variable as an F/1O Cepheid.\newline
\noindent{\it 5895841249526120064}: This star has a total amplitude of only $\sim 0.13$\,mag in the Gaia $G$-band light curve, and no sign of a secondary periodicity. The slight asymmetry of the light curve hints at fundamental mode pulsation, as either a BL~Her or a classical Cepheid variable.\newline
\noindent{\it 6117651360865355136}: This star is the RV~Tau type variable V820~Cen. It has only 19 epochs in the Gaia $G$ band, leading to its misclassification.\newline
\noindent{\it 6394890542044709888}: This star is ASASSN-V J031346.98+632052.8, classified as a small-amplitude classical Cepheid (DCEPS) by \cite{Jayasinghe2018}. There is no sign of additional periodicities in either the Gaia or the ASASSN light curves. We do note that its low Galactic latitude ($-49.795\deg$) and faintness would lead to a large inferred distance from the Galactic disk, if the classification as a classical Cepheid would hold true for this variable, therefore it is most probably some other kind of variable.\newline
\noindent{\it 6710614339593008384}: This variable is the RRab star SSS\_J185900.5-462532 \citep{Torrealba2015}, with a pulsation period of $\sim0.53518$\,days. The low number of Gaia epochs (15) has led to the misclassification of this variable.\newline

\FloatBarrier

\section{Galactic metallicity gradient}
\label{grad}

\subsection{Galactocentric distance}

\par
\begin{figure}[!t]
\centering
\includegraphics[width=9cm]{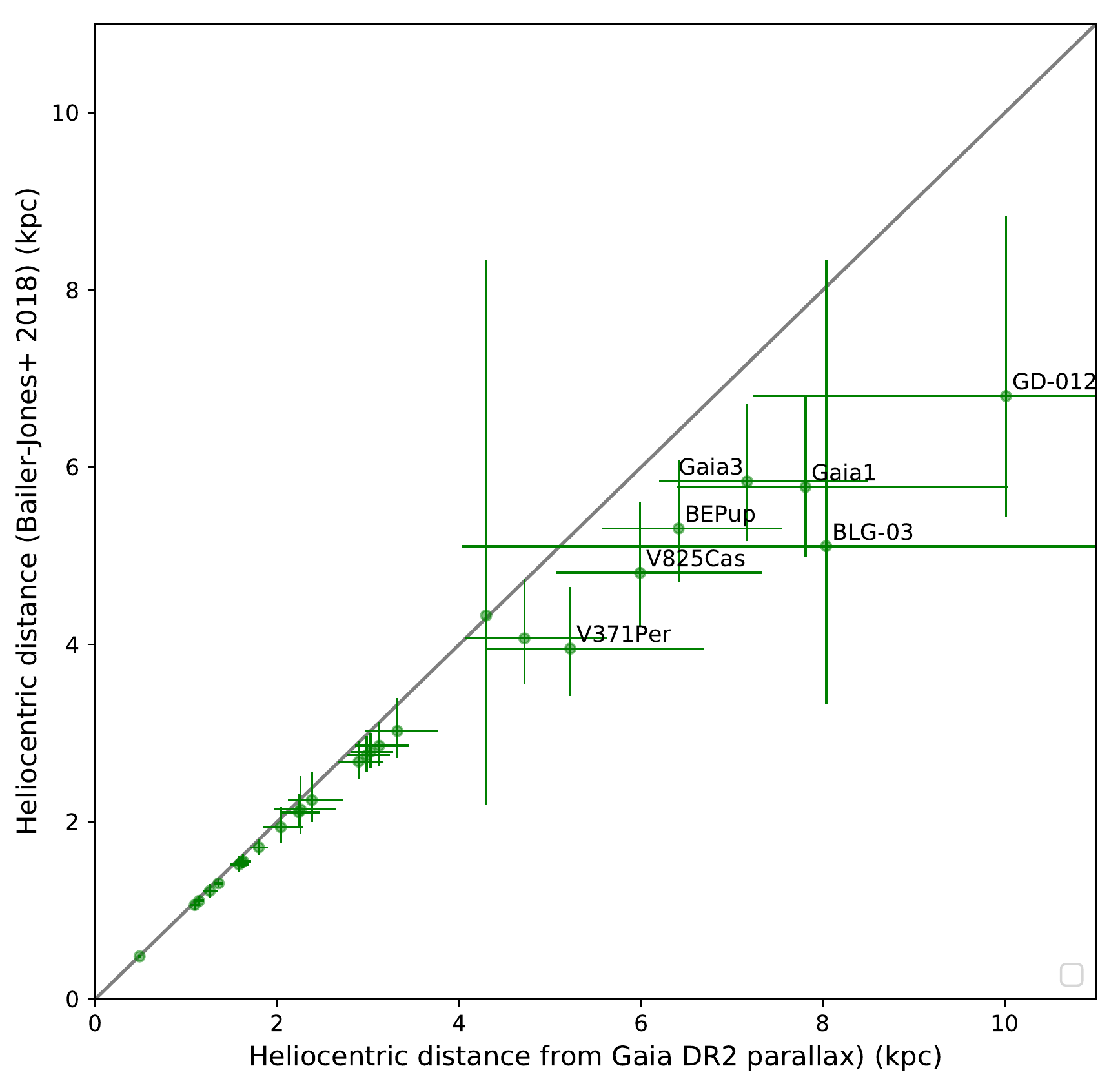}
\caption{Comparison of the heliocentric distance computed by inverting the Gaia DR2 parallax to the heliocentric distance derived by \cite{Bailer-Jones2018} using a geometrical distance prior.}
\label{comp_helio_dist}
\end{figure}

 For all but two stars in our sample (that have negative parallaxes), we can derive the heliocentric (and, in turn, Galactocentric) distance by inverting the parallax value provided by Gaia DR2. In Fig.~\ref{comp_helio_dist} we compare those distances to the distances computed by \cite{Bailer-Jones2018} using a purely geometrical distance prior relying on a model of the Milky Way. 

\par Both sets of distances are in excellent agreement as long as the stars are located within $\sim$4--5 kpc from the Sun, a distance beyond which the distance estimates start to diverge. This is presumably due to the combined effect of the decreasing accuracy of the Gaia parallaxes at large distances and of the uncertainties of the Galaxy model from which the priors are determined.

\par
\begin{figure}[!t]
\centering
\includegraphics[width=9cm]{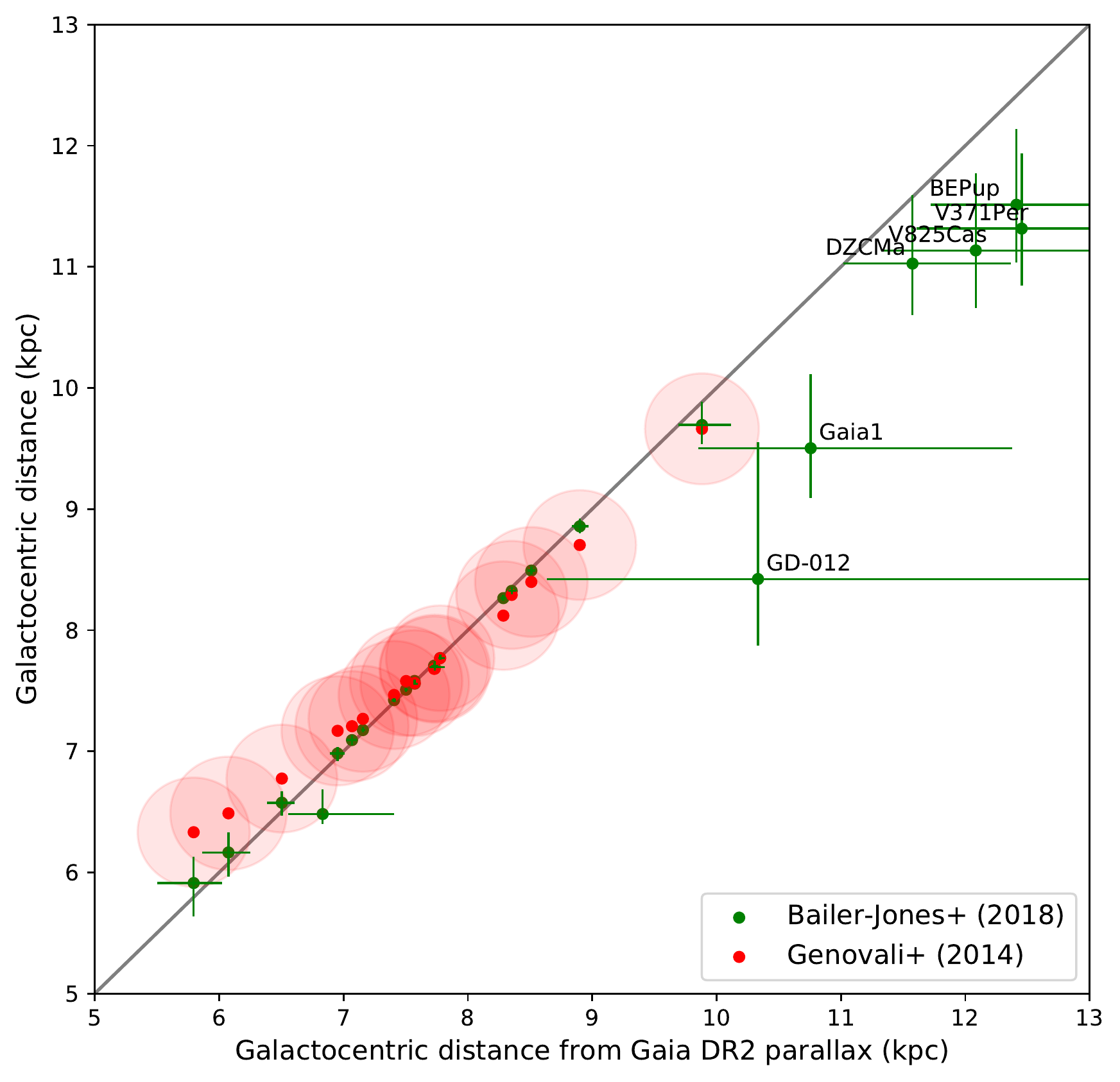}
\caption{Comparison of the Galactocentric distances computed using Gaia DR2 parallaxes only, or combining them with a geometrical distance prior \citep{Bailer-Jones2018}. Galactocentric distances derived from near-infrared photometry \citep{Inno2013,Genovali2014} and their associated error-bars are shown in red. For comparison purposes, all the distances have been computed assuming that the Sun is located at 7.94 kpc from the Galactic center \citep{Matsunaga2013}.}
\label{comp_galac_dist}
\end{figure}

 Photometric distances \citep{Genovali2014} based on reddening-free period-Wesenheit\footnote{Wesenheit indices are pseudo-magnitudes related to apparent magnitudes, but minimally affected by uncertainties on reddening by construction, \citep[see][]{Madore1982}. They were computed using the reddening law of \citet{Cardelli1989}.} relations in the near-infrared \citep{Inno2013} are also available for some of the stars in our sample, however the accuracy of those distances is not homogeneous: in the absence of near-infrared time-series data, some of them have been derived using a single-point value taken from 2MASS together with light curve templates. Photometric Galactocentric distances are nevertheless in excellent agreement with Gaia parallax-based distances as shown in Fig.~\ref{comp_galac_dist}. The small divergence observed in the inner part of the disk is presumably related to increasing uncertainties on the reddening law and individual reddening values at large distances in the disk. Since they are available for all but two Milky Way F/1O Cepheids, we used distances directly derived by inverting the Gaia DR2 parallax in the rest of the paper.

\begin{table}[!ht]
\footnotesize
\caption{Galactocentric distances derived from Gaia DR2 parallaxes \citep{GaiaCollaboration2018}, Gaia DR2 parallaxes combined with a geometrical prior \citep{Bailer-Jones2018} or period-Wesenheit relations in the near-infrared \citep{Inno2013,Genovali2014}. Distances have been computed assuming a Galactocentric distance of 7.94 kpc for the Sun.}        
\label{distances}
\centering
\begin{tabular}{l r@{\hskip6pt}l r r r} 
\hline\hline
\multicolumn{3}{c}{Star} & \multicolumn1c{R$_{g}$} & \multicolumn1c{R$_{g}$} & \multicolumn1c{R$_{g}$} \\
\multicolumn{3}{c}{    } & \multicolumn1c{kpc} &   \multicolumn1c{kpc} & \multicolumn1c{kpc} \\
\multicolumn{3}{c}{    } & \multicolumn1c{Gaia parallax} & \multicolumn1c{Gaia parallax} & \multicolumn1c{PW (NIR)}\\
\multicolumn{3}{c}{    } &               & \multicolumn1c{with prior}  &         \\
\hline
& V371 & Per	                 & 12.454 & 11.315 &       \\	     
& TU & Cas   	                 &  8.510 &  8.492 & 8.398 \\	     
& V825 & Cas   	                 & 12.084 & 11.135 &       \\	     
& AS & Cas     	                 &  9.883 &  9.969 & 9.662 \\ 
& Gaia & 1    	                 & 10.757 &  9.502 &       \\	     
& DZ & CMa    	                 & 11.575 & 11.026 &       \\	     
\multicolumn3l{OGLE-BLG-CEP-21}  &  3.652 &  3.622 &       \\    
\multicolumn3l{OGLE-BLG-CEP-03}  &  0.611 &  2.878 &       \\
& V458 & Sct                     &  6.075 &  6.165 & 6.488 \\
& V367 & Sct                     &  5.795 &  5.913 & 6.332 \\
& EW & Sct                       &  7.503 &  7.508 & 7.580 \\
& BQ & Ser                       &  6.952 &  6.982 & 7.169 \\
\multicolumn3l{OGLE-GD-CEP-0009} &        &  9.559 &       \\
& EY & Car                       &  7.573 &  7.559 & 7.563 \\
& GZ & Car                       &  7.728 &  7.707 & 7.681 \\
& V701 & Car                     &  7.777 &  7.769 & 7.769 \\
& BK & Cen                       &  7.156 &  7.176 & 7.268 \\
& UZ & Cen                       &  7.406 &  7.423 & 7.465 \\
\multicolumn3l{OGLE-GD-CEP-0012} & 10.333 &  8.422 &       \\ 
& Y & Car                        &  7.734 &  7.694 & 7.682 \\
& Gaia & 2                       &  7.571 &  7.581 &       \\
& AX & Vel                       &  8.285 &  8.264 & 8.120 \\
& AP & Vel                       &  8.352 &  8.324 & 8.291 \\
& BE & Pup                       & 12.411 & 11.512 &       \\ 
& VX & Pup                       &  8.900 &  8.857 & 8.703 \\
& U & TrA                        &  7.068 &  7.093 & 7.207 \\
& Gaia & 3                       &  6.832 &  6.482 &       \\
\multicolumn3l{OGLE-GD-CEP-0016} &        &  8.563 &       \\
& V1210 & Cen                    &  6.504 &  6.576 & 6.775 \\
& V901 & Mon                     &  8.142 &  8.141 &       \\
\hline
\end{tabular}
\end{table}

\subsection{Spatial distribution of the F/1O Cepheids in our sample and sample selection}

\par
\begin{figure}[!h]
\centering
\includegraphics[width=9cm]{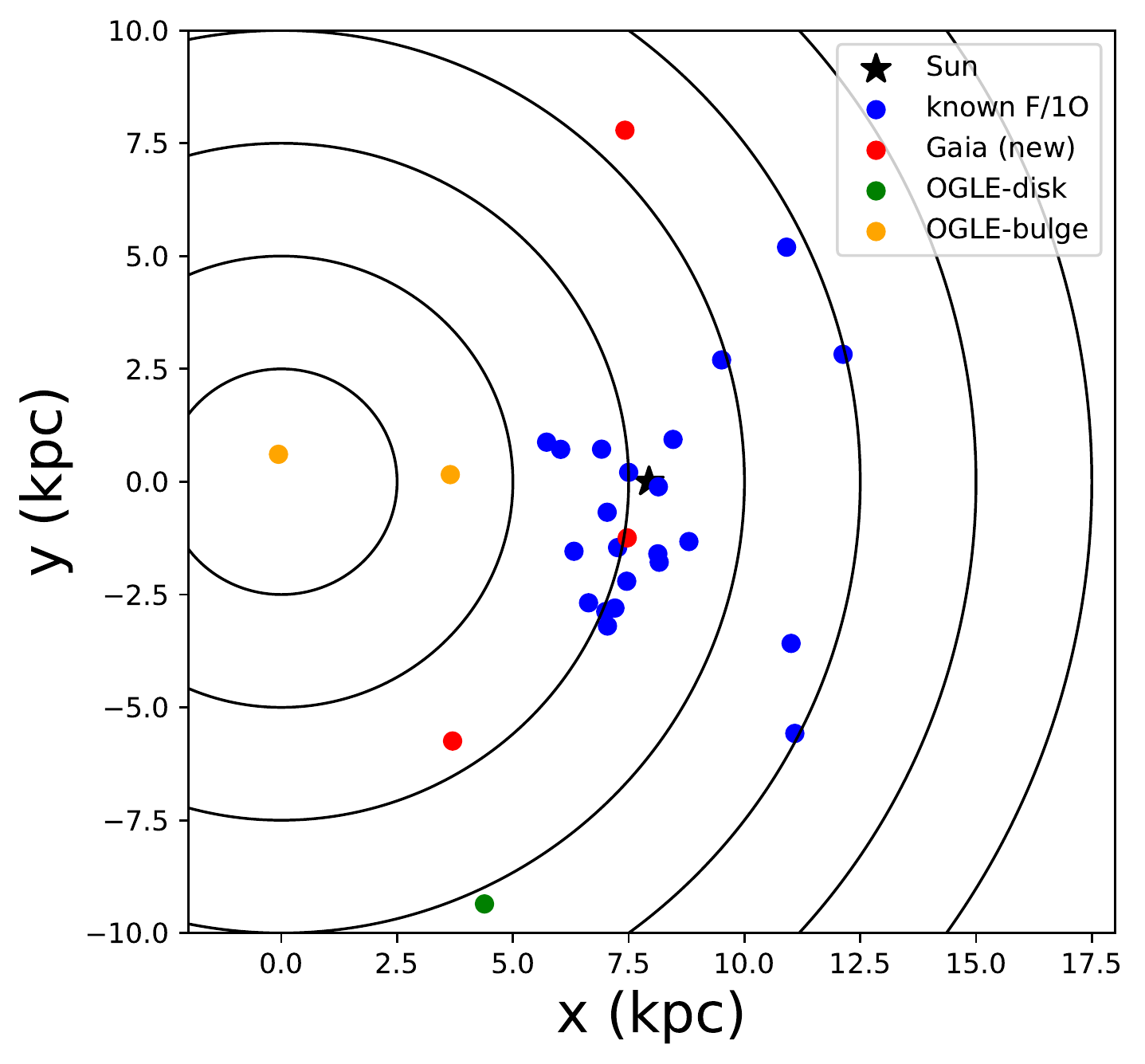}
\caption{Spatial distribution of the F/1O Cepheids in the Milky Way. The currently known F/1O Cepheids are shown in blue, and in yellow/green for the bulge/disk Cepheids in the OGLE sample of \cite{Soszynski2011} and \cite{Pietrukowicz2013}, respectively. The new F/1O Cepheids in Gaia DR2 are shown in red.}
\label{Spatial_dist}
\end{figure}

 Fig.~\ref{Spatial_dist} shows the spatial distribution of the F/1O Cepheids in the Milky Way. The location of the stars has been computed using as heliocentric distance the inverse of the Gaia DR2 parallaxes, and two of the Cepheids in the OGLE disk sample are therefore missing because they have negative parallaxes in Gaia DR2.

\par For the derivation of the gradient, we did not include the OGLE Cepheids in the disk or towards the bulge. As already mentioned, 
two of the disk ones have negative parallaxes in Gaia DR2 and no homogeneous distances can be determined for these stars. Moreover, the exact location of the OGLE Cepheids towards the bulge remains quite uncertain, one of them being even placed in the flared outer disk at the far side of our Galaxy by \cite{Feast2014}.

\par Furthermore, all but one of the OGLE F/1O Cepheids have fundamental periods shorter than two days, while the relation of \cite{Kovtyukh2016} has been calibrated with F/1O Cepheids with fundamental periods spanning a [2d--6d] range. Metallicities derived by applying this formula to shorter period Cepheids might be inaccurate, as it seems to be the case for the  F/1O Cepheid Gaia3, with a fundamental period slightly larger than one day.   

\subsection{Computation of the gradient: method}
\label{bootstrap}

\par Our sample of F/1O Cepheids adopted to derive the Milky Way metallicity gradient therefore contains 25 stars, including the 3 new F/1O Cepheids discovered by Gaia. The determination of the gradient requires both distances and metallicities for the stars in our sample. Here we use a simple bootstrap method to derive the slope and the intercept of the gradient and their standard deviation.

\par The first step in our analysis consists in estimating the error on the metallicity derived using the \cite{Kovtyukh2016} relation. To achieve this goal, we have generated 10000 relations between [Fe/H], $log(P_{0}$) and $P_{1}/P_{0}$ in a multivariate normal (Gaussian) distribution, for which we use as input parameters the coefficients and the covariance matrix of the function fitted by \cite{Kovtyukh2016}. For a given Cepheid, the standard deviation of the 10000 values of [Fe/H] generated is adopted as the error on the metallicity. Having estimated the errors on the metallicity, we can now draw random metallicities assuming a normal distribution around the value given by the \cite{Kovtyukh2016} formula.

\par For the distances, we draw random parallaxes assuming a normal distribution of the error on the parallax, where the value reported by Gaia DR2 is considered as the standard deviation of the distribution. This parallax is converted into heliocentric and then Galactocentric distance $R_{G}$. We assume a Galactocentric distance of 7.94 kpc for the Sun, in order to enable a direct comparison with the gradient obtained by \cite{Genovali2014} for numerous Cepheids pulsating in various modes. We investigated the influence of the adopted Solar Galactocentric distance and results are given in Table~\ref{GC}.

\par We realize 10000 drawings of distances and metallicities for the 25 stars in our sample, which means that for a given population we draw 25 (distance, metallicity) pairs, and we repeat the operation 10000 times. We then fit a linear gradient to each of these populations and obtain 10000 slopes and intercepts. Their means and standard deviations give us the slope and intercept of the Galactic metallicity gradient and associated errors.

\subsection{Computation of the gradient: alternative method}
\label{TLS}

\par We also computed the Milky Way metallicity gradient using a total least squares regression \citep{Hogg2010}, as implemented in the \texttt{astroML} python package \citep{VanderPlas2012,Ivezic2014}. With this method, observational errors on both variables $R_{G}$ and [Fe/H] are taken into account. For a given star, the standard deviation of the 10000 realizations of $R_{G}$ computed for the bootstrap method is adopted as the uncertainty on $R_{G}$. In this case, for every Cepheid, the two quantities  ($R_{G}$, [Fe/H]) are independently determined, therefore their estimates are not correlated. We note however, that the distributions of the uncertainties  (especially on $R_{G}$) are not exactly Gaussian.

\subsection{Results}

\par
\begin{figure}[t]
\centering
\includegraphics[width=9cm]{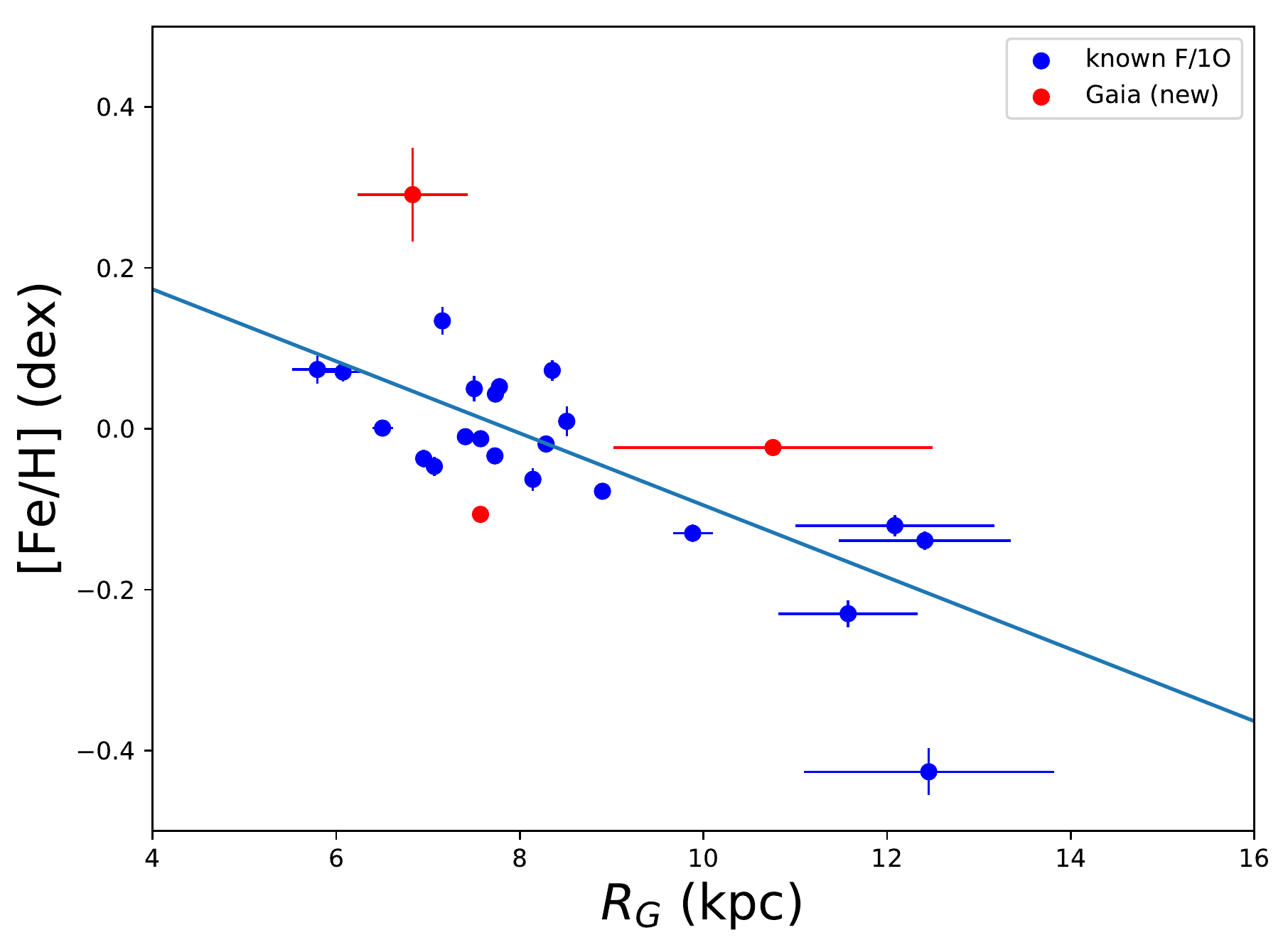}
\caption{Metallicity gradient in the disk where distances and metallicities are the nominal values derived from the Gaia DR2 parallaxes and the \cite{Kovtyukh2016} relation, respectively. The color coding is the same as in Fig.~\ref{Spatial_dist}. The slope and the intercept of the gradient have been computed using a bootstrap method (see Sect.~\ref{bootstrap}).}
\label{Gradient}
\end{figure}

\begin{figure}[t]
\centering
\includegraphics[width=9cm]{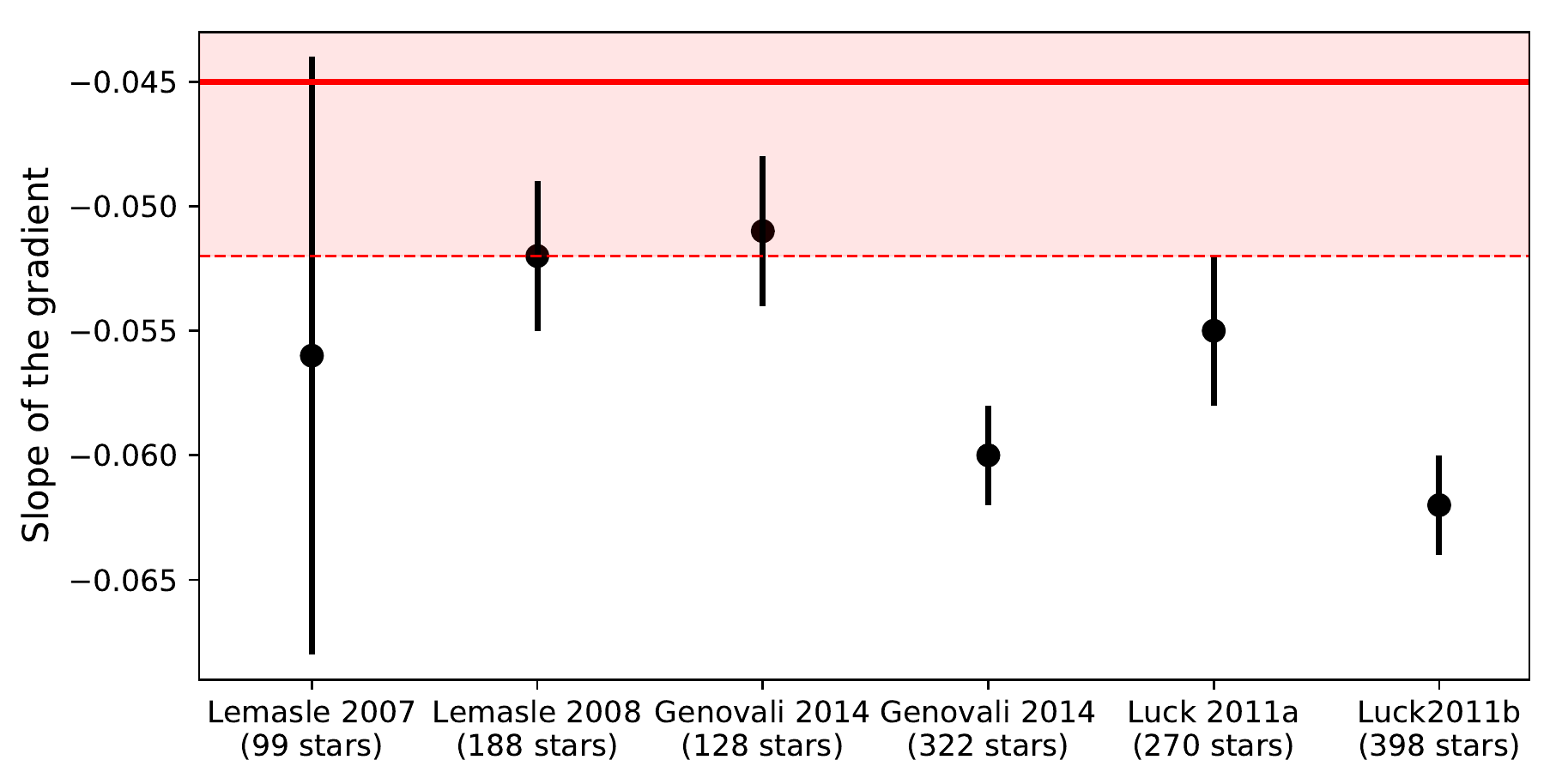}
\caption{Comparison of the slopes of the present-day [Fe/H] gradient derived from Cepheids in different studies \citep{Lemasle2007,Lemasle2008,Luck2011a,Luck2011b,Genovali2014}. Slopes and error bars are shown in black while the red line/red area represent the result of the current study using 24 F/1O Cepheids (Bootstrap method, R$_{g\odot}$=7.94~kpc).}
\label{comparison}
\end{figure}

\begin{figure}
\centering
\includegraphics[width=9cm]{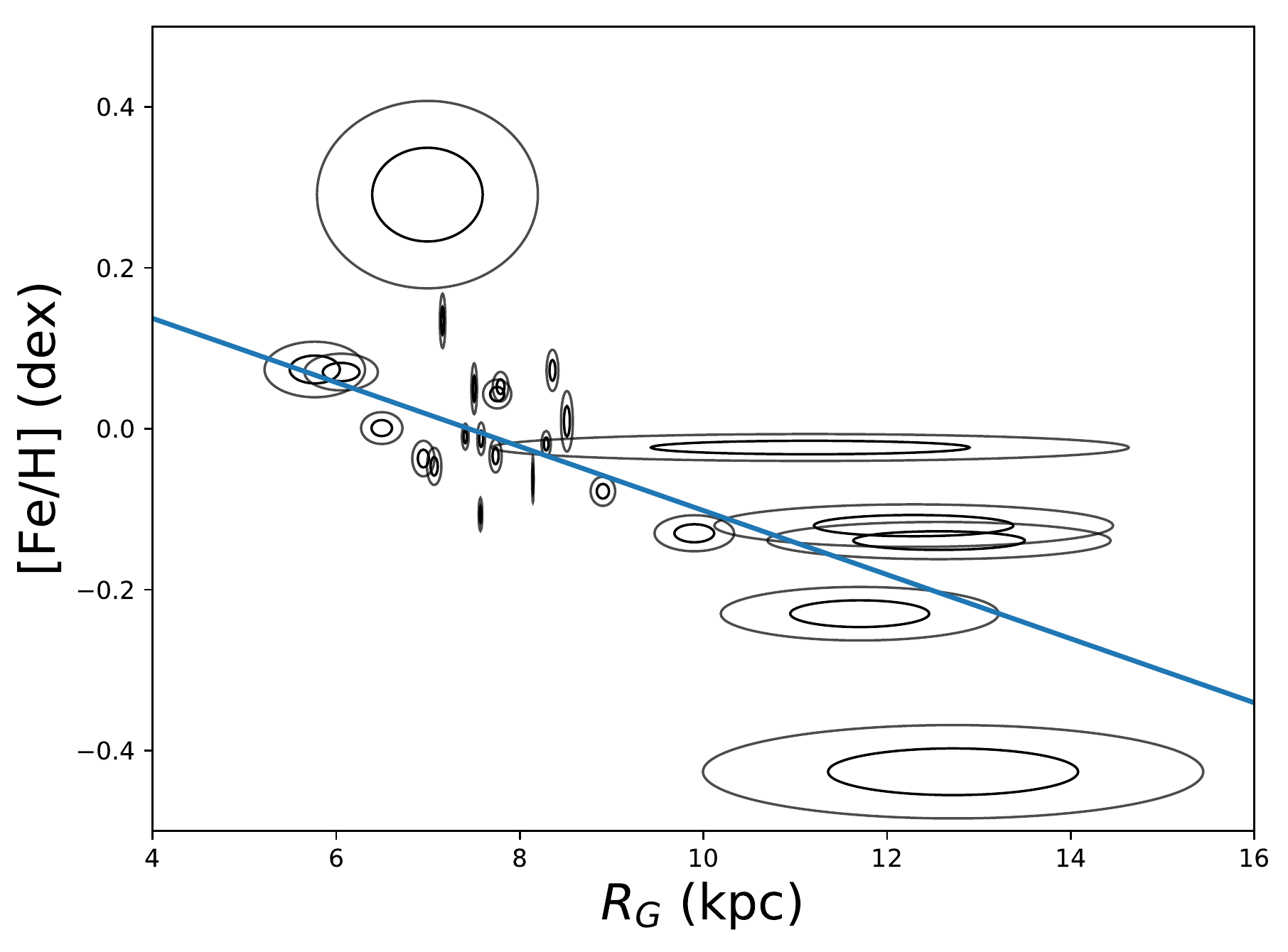}
\caption{Metallicity gradient in the disk where distances and metallicities are the nominal values derived from the Gaia DR2 parallaxes and the \cite{Kovtyukh2016} relation, respectively. The slope and the intercept of the gradient have been computed using a total least square method (see Sect.~\ref{TLS}). The ellipses trace the 1--$\sigma$ and 2--$\sigma$ likelihood contours.}
\label{Gradient_TLS}
\end{figure}

 With the bootstrap method, we obtain a slope of --0.0447$\pm$0.0066 dex/kpc for the gradient (see Fig.~\ref{Gradient}), in good agreement with the values obtained by e.g., \cite{Lemasle2007} (--0.061$\pm$0.19 dex/kpc), \cite{Lemasle2008} (--0.052$\pm$0.03 dex/kpc) or the own sample of \cite{Genovali2014} (--0.052$\pm$0.004 dex/kpc). This slope is slightly lower than the slope derived by \cite{Genovali2014} (--0.055$\pm$0.002 dex/kpc) combining their data with literature values adjusted to a common metallicity scale. or by \cite{Luck2011a,Luck2011b} who found slopes of -0.055$\pm$0.003 and -0.062$\pm$0.002 dex/kpc, respectively. A comparison of the slopes mentioned here can also be found in Fig.~\ref{comparison}. It is also worth mentioning that the range in Galactocentric distances covered by the different estimates of the metallicity gradient changes among the different samples. The agreement remains reasonable with the total least squares method (Fig.~\ref{Gradient_TLS}), which gives a (shallower) slope of --0.040$\pm$0.002 dex/kpc.
\FloatBarrier

\par As can be seen in Fig.~\ref{Gradient_TLS}, the uncertainties on both [Fe/H] and the Galactocentric distance are small for the nearby Cepheids, because their parallaxes could be determined with a good accuracy. In contrast, the less accurate parallaxes for distant Cepheids lead to stars that are at best only marginally consistent with the computed slope. Better parallaxes at large distances from Gaia DR2 and a larger sample of F/1O Cepheids in the outer disk would help to better constrain the metallicity gradient in this region.

\par Only one star has a very large uncertainty on metallicity. It is one of the newly discovered F/1O Cepheids, and the uncertainty on the derived metallicity may be related to the fact that its fundamental period ($\approx$ 1d) falls out of the range of periods ([2d--6d]) for which the \cite{Kovtyukh2016} relation was calibrated. Despite its also quite large uncertainty on the distance, it is unlikely that this Cepheid would follow the general trend if the error bars on the measurements would be reduced. This could indicate that the star was born in a slightly different environment than the other Cepheids, and indeed Fig.~\ref{Spatial_dist} shows that it is located in a different region of the disk. Alternatively it could prove that the current relation between $P_{1}/P_{0}$ and [Fe/H] does not hold anymore at short periods.

\par The quoted results were obtained assuming that the Sun is located at 7.94 kpc from the Galactic center, in order to allow for direct comparisons with the values obtained by \cite{Genovali2014}. These authors adopted this distance as it was derived from similar tracers (classical Cepheids) discovered in the Galactic nuclear bulge by \cite{Matsunaga2013}. We checked however that both with the bootstrap and total least squares methods, the slope and intercept of the gradient are only marginally affected by the choice of other values (8.0 kpc, 8.3 kpc) for R$_{g\odot}$. Results are tabulated in Table~\ref{GC}.

\begin{table*}
\caption{Influence of the adopted distance to the Galactic Center on the slope and intercept of the metallicity gradient.}        
\label{GC}
\centering
\begin{tabular}{c c | c c | c c} 
\hline\hline
 & & \multicolumn{2}{c}{Bootstrap method} & \multicolumn{2}{c}{Total least squares method} \\
 \hline
R$_{g}$ & Reference & slope & intercept & slope & intercept\\
kpc     & & dex/kpc &  dex & dex/kpc &  dex\\
\hline
7.94 & {\tiny \cite{Groenewegen2008b,Matsunaga2013}} & -0.0447$\pm$0.0066 & 0.3522$\pm$0.0528 & -0.0398$\pm$0.0024 & 0.2963$\pm$0.0181 \\
8.00 & {\tiny \cite{Reid1993,Camarillo2018}} & -0.0449$\pm$0.0066 & 0.3558$\pm$0.0532 & -0.0398$\pm$0.0025 & 0.2988$\pm$0.0181 \\
8.30 & {\tiny \cite{deGrijs2016,Majaess2018}} & -0.0455$\pm$0.0066 & 0.3737$\pm$0.0550 & -0.0399$\pm$0.0025 & 0.3112$\pm$0.0189 \\
\hline
\end{tabular}
\end{table*}


\par In distant systems, metallicity gradients derived from observations usually refer to the [O/H] gradient measured in HII regions or planetary nebulae. In addition to the intrinsic uncertainties related to the abundance determination in such tracers, the comparison with iron gradients derived from Cepheids requires the transformation of the measured [O/H] into $Z$ or [Fe/H], which may vary from system to system or even within the observed system. 
With these caveats in mind, it is interesting to note that \cite{Beaulieu2006} found a good agreement between their gradients and those derived from HII regions \citep[e.g.,][]{Garnett1997}, B supergiants \citep[e.g.,][]{Urbaneja2005} or planetary nebulae \citep[e.g.,][]{Magrini2004} in M33. In M31, \cite{Lee2013} report a good agreement between their double-mode Cepheids gradient and the gradient derived from HII regions \citep{Sanders2012,Zurita2012}, but the agreement is surprisingly even better with the slope derived by \cite{Kwitter2012} from planetary nebulae. Our study validates the use of metallicity gradients derived from F/1O Cepheid period ratios through direct comparison to those obtained using full spectroscopic analyses of related objects, the single-mode classical Cepheids.

\section{Conclusions}
\label{Conc}

\par We have identified only 3 new Galactic F/1O Cepheids within the 30 Gaia DR2 candidates. After inspection of their light curves and with the help of the literature, we propose an alternative classification for the remaining stars. With a larger number of individual measurements and hence better populated light curves, it is very likely that the number of genuine F/1O Cepheids will increase in the future Gaia releases.
\par Thanks to the accurate Gaia DR2 parallaxes, we have derived Galactocentric distances for the almost entire sample of known Galactic F/1O Cepheids. Taking advantage of the metallicity dependence of the $P_{1}/P_{0}$ ratio, we have derived the present-day Milky Way metallicity gradient in the thin disk. The slope of --0.045$\pm$0.007 dex/kpc is in good agreement with the gradient determined using Cepheids pulsating in various modes, which validates the use of F/1O Cepheids to derive metallicity gradients. We recommend high resolution  spectroscopic follow-up of short-period F/1O Cepheids in order to extend the period range over which the \cite{Kovtyukh2016} relation is applicable. 

\begin{acknowledgements}
We thank the anonymous referee for her/his suggestions that helped to improve  the quality of the paper. B.L., G.H., L.I. \& E.K.G. acknowledge support from the Sonderforschungsbereich SFB 881 ‘The Milky Way System’ (sub-projects A3, A5) of the German Research Foundation (DFG). Additional support for M.C. and G.H. for this project is provided by Fondecyt through grant \#1171273; the Ministry for the Economy, Development, and Tourism's Millennium Science Initiative through grant IC\,120009, awarded to the Millennium Institute of Astrophysics
(MAS); by Proyecto Basal PFB-06/2007; and by CONICYT's PCI program through grant DPI20140066.
G.H. acknowledges additional support from the Graduate Student Exchange Fellowship Program between the Institute  of Astrophysics of the Pontificia Universidad Cat\'olica de Chile and the Zentrum f\"ur Astronomie der Universit\"at Heidelberg, funded by the Heidelberg Center in Santiago de Chile and the Deutscher Akademischer Austauschdienst (DAAD), and by the CONICYT-PCHA/Doctorado Nacional grant 2014-63140099.
A.K. acknowledges the National Research Foundation of South Africa and the Russian Science Foundation (project no. 14-50-00043). This work has made use of data from the European Space Agency (ESA) mission
{\it Gaia} (\url{https://www.cosmos.esa.int/gaia}), processed by the {\it Gaia}
Data Processing and Analysis Consortium (DPAC,
\url{https://www.cosmos.esa.int/web/gaia/dpac/consortium}). Funding for the DPAC
has been provided by national institutions, in particular the institutions
participating in the {\it Gaia} Multilateral Agreement.
This research has made use of the SIMBAD database, operated at CDS, Strasbourg, France \citep{Wenger2000}. This research has made use of the International Variable Star Index (VSX) database, operated at AAVSO, Cambridge, Massachusetts, USA. This research has made use of the \texttt{TOPCAT}\footnote{\url{http://www.star.bristol.ac.uk/~mbt/topcat/}} software \citep{Taylor2005}.
This research has made use of the \texttt{numpy}, \texttt{scipy}, \texttt{matplotlib} \citep{Hunter2007}, \texttt{astropy} \citep{TheAstropyCollaboration2018},  \texttt{iPython}, 
and \texttt{astroML} \citep{VanderPlas2012,Ivezic2014}
python packages.
\end{acknowledgements}


\bibliographystyle{aa} 
\bibliography{lemasle.bib} 

\end{document}